\pdfoutput=1

\documentclass[twocolumn,showpacs,preprintnumbers,amsmath,amssymb]{revtex4-1}
\usepackage{dcolumn}
\usepackage{bm}

\newcommand{\be}{\begin{equation}}
\newcommand{\ee}{\end{equation}}
\newcommand{\bea}{\begin{eqnarray}}
\newcommand{\eea}{\end{eqnarray}}

\global\long\def\myvec#1{ \ensuremath{\mathbf{\underline{{#1}}}}}

\usepackage[dvips]{color}
\usepackage{epsfig}
\usepackage{ulem}
\usepackage{amsmath}

\begin{document}

\preprint{APS/unknown}

\title{Interfacial coupling across a modified interface studied with ferromagnetic resonance}

\author{R.  Magaraggia}
\email{rhet.magaraggia@gmail.com}
\author{S. McIntyre}%
\author {K. Kennewell}%
\affiliation{School of Physics, University of Western Australia, 35 Stirling Highway, Crawley, Western Australia 6009, Australia }

\author{R. L. Stamps}
\affiliation{School of Physics, University of Western Australia, 35 Stirling Highway, Crawley, Western Australia 6009, Australia }
\affiliation{SUPA, School of Physics and Astronomy, University of Glasgow, Glasgow G12 8QQ, United Kingdom}

\author{M. Ali}
\author{D. Greig}
\author{B. J. Hickey}
\author{C. H. Marrows}
\affiliation{School of Physics and Astronomy, E. C. Stoner Laboratory, University of Leeds, Leeds LS2 9JT, United Kingdom }%

\date{\today}

%
%

\begin{abstract}
Using spin waves we directly probe the interface of an exchange biased
Ni$_{80}$Fe$_{20}$/Ir$_{25}$Mn$_{75}$ film which has been modified
by the presence of an Au dusting layer. Combining this experimental
data with a discretised simulation model, parameters relating to interface
exchange coupling and modification of interface magnetisation are
determined. Exchange coupling is found to be relatively uniform as
gold thickness is increased, and undergoes a sudden drop at 1.5$\textrm{\AA}$
of gold. Interface magnetisation decreases as a function of the gold dusting thickness.  Antiparallel alignment of the ferromagnet and antiferromagnet supress the interface magnetisation compared to when they are in parallel alignment.
These findings imply that the interface region has specific magnetisation
states which depend on the ferromagnet orientation.
\end{abstract}


\maketitle

\section{Introduction}

Fine scale control over growth of magnetic interfaces has made the
tailoring of bulk magnetic properties though exchange bias possible\cite{PhysRevB.34.1689,PhysRevLett.89.077201,Nature.423.2003}. In most cases it is not at all clear what occurs at the interface
region and the role played by various factors such as roughness\cite{NatMat.5.128.2006},
intermixing\cite{PhysRevB.71.054411}, coupling direction\cite{PhysRevB.59.6984,PhysRevB.65.134436,PhysRevLett.99.037204}
and defects\cite{PhysRevB.66.014431,PhysRevB.79.134415}. Recently,
many groups have begun to make detailed observations of this region
with techniques able to probe buried interfaces such as
neutron scattering\cite{PhysRevLett.99.037204,PhysRevLett.95.047201,ball:583},
XMCD\cite{bruck:126402,PhysRevLett.95.047201,PhysRevLett.91.017203,PhysRevB.81.212404,PhysRevB.72.024421} and
electron M\"ossbauer\cite{PhysRevB.70.224414}. Ferromagnetic resonance
\cite{PhysRevB.83.054405,screening-published,PhysRevB.38.6847,PhysRevB.58.8605,Kittel_Formula,kuanr:07C107,PhysRevB.70.094420,PhysRevB.65.064410} can also be used to study
magnetic interfaces through shifts in the spin wave mode resonance conditions.  Sufficiently thick films allow multiple standing spin wave modes to be observed. Their standing wave profiles are distorted
by magnetic interactions at the FM/AFM interface. It has been shown
that by interpreting these standing wave modes with a suitable model,
information about interface coupling and interface magnetisation may be
obtained\cite{PhysRevB.54.4159,PhysRevB.83.054405,Classic_Kittel_FMR,Spin_Wave_pinning,PhysRevB.19.4575,PhysRevB.72.014463,PhysRevB.69.134426}. This is possible because interface pinning affects higher order mode frequencies more strongly then lower order mode frequencies due to the quadratic spin wave dispersion relationship \cite{Kittel_Formula}. 

In order to uniquely identify the role of the interface in exchange
bias, many experiments have been performed which directly modify
the interface region either through ion-bombardment\cite{Fassbender2008579,mewes:1057,PhysRevB.63.060409,juraszek:6896,bomb-2010-unpublished}
or direct doping using an impurity layer\cite{PhysRevB.77.134401,PhysRevB.78.205430,eurolett.54.2.2002}.
Here we study Permalloy/IrMn bi-layers that have a partial Au layer
between the ferromagnetic and antiferromagnetic layers. This partial
gold layer is not thick enough to form a continuous layer, and is
characterised though an nominal thickness.  Gold is chosen because it locally blocks exchange coupling, as opposed to other materials such as iron which may enhance local exchange coupling\cite{PhysRevB.77.134401}.
This partial or {}``dusting'' layer at the interface allows a detailed
examination of how the FM/AFM interaction changes\cite{PhysRevB.77.134401}
as a result of interface disruption.

Ferromagnetic resonance experiments were carried out at microwave frequencies
to probe the magnetic ground state of the system. These resonances
are calculated with an atomistic model which incorporates dipolar and
exchange coupling as well as allowing a detailed adjustment of the
layer by layer parameters. Measures of effective interface coupling
are extracted which replicate the observed resonance frequencies.
The simplest fit, with the fewest parameters, is to assume that the
interface magnetisation must take on different values depending on
the ferromagnet orientation with respect to the antiferromagnet. The
gold dusting layer is found to disrupt the interface magnetisation,
and interface coupling is dramatically reduced for nominal thicknesses
less then a continuous monolayer.

\section{Experiment and Characterisation}

Magnetic multilayer specimens consisting of Ta(50 $\textrm{\AA}$)/
Ni$_{80}$Fe$_{20}$ (605 $\textrm{\AA}$)/Au dusting (t $\textrm{\AA}$)/Ir$_{25}$Mn$_{75}$
( 60 $\textrm{\AA}$)/ Ta(50 $\textrm{\AA}$) and Ta(50 $\textrm{\AA}$)/
Ni$_{80}$Fe$_{20}$ (t $\textrm{\AA}$)/Ir$_{25}$Mn$_{75}$ ( 60
$\textrm{\AA}$)/ Ta(50 $\textrm{\AA}$) were sequentially deposited
onto Si(001) substrates by dc-magnetron sputtering at an argon working
pressure of 2.5 mTorr. Typical deposition rates were 2\textendash{}2.5
 $\textrm{\AA}$ s$^{-1}$, which were determined by measuring the thickness of calibration
films by low-angle x-ray reflectometry. Film roughness was in the
order of 3-4 $\textrm{\AA}$, also determined by low-angle x-ray reflectometry. An in-plane field of 15.9 kA m$^{-1}$ was applied
during the growth to induce a macroscopic uniaxial anisotropy in the
Ni$_{80}$Fe$_{20}$ layer in a defined direction. The base pressure
prior to the deposition was of the order of 1$\times$10$^{-8}$ Torr
and the samples were deposited at ambient temperature. 

The IrMn layers were deposited from an alloy target. To facilitate
the growth of face-centred-cubic (fcc) (111) orientation of IrMn, a buffer
underlayer of Ta was employed. X-ray diffraction revealed that such
samples were predominantly fcc with a (111) texture. We did not detect
any changes in texture in a representative selection of doped samples
measured by this technique, presumably since the Au dusting layer
was so thin.  Furthermore, the fcc structure of Au should prevent disruption of the crystal structure in subsequent layers. No post annealing steps were required, since the exchange bias pinning
direction was set by a 15.9 kA m$^{-1}$ in-plane forming field applied to the
sample during the deposition of all the layers in this top spin-valve
configuration. 

The IrMn layer thickness of 60 $\textrm{\AA}$ was chosen such that
any slight changes in IrMn layer thickness itself did not alter the
exchange bias. Atomic level disorder at the interface was
achieved by depositing a $\delta$ layer dusting of Au \cite{PhysRevB.77.134401}.
The Au dusting layer was varied from 0 to 1.5 $\textrm{\AA}$,
and is discontinuous for layer thickness below
8 $\textrm{\AA}$, given that the lattice spacing for Au
is 4.08 $\textrm{\AA}$. 

In-plane and out-of-plane FMR magnetometry was used to extract $\mu_{0}M_{S}$
from resonance frequency data\cite{Kittel_Formula}.
In-plane FMR magnetometry along the easy axis of a Ni$_{80}$Fe$_{20}$ sample with
no IrMn gave a saturation magnetisation $\mu_{0}M_{S}$ of 0.80$\pm$0.05 T,
a gyromagnetic ratio of 2.8$\times$10$^{10}$ Hz T$^{-1}$ and in
plane bulk anisotropy $\mu_{0}H_{A}$ of less then 0.0007 T.
Further magnetometry was performed using the magneto-optical Kerr
effect (MOKE). A 635 nm diode laser, rated at 5 mW, was used to illuminate
the sample. A differential amplifier was used to analyse polarisation
rotation. 

MOKE measurements were carried out along the sample easy axis by sweeping the field at 10 Hz, in order to reduce noise. Exchange bias is determined from MOKE measurements by taking half
the difference between the positive and negative coercive field points
on the corresponding hysteresis loop.

Broadband FMR measurements were carried out using a 20 GHz vector network
analyser to excite and detect resonance measurements. A 0.3 mm wide
microstrip line, which was connected to 50 $\Omega$ high frequency
co-axial cables, acts as a microwave antenna and drives resonance
in the sample, which is placed on top and is in direct contact with
the microstrip. A Kepco powered electromagnet is used to supply the
in-plane magnetic field. This magnetic field is swept and the S21(H)
parameters are measured at a fixed microwave frequency as in \cite{screening-published,PhysRevB.83.054405}. The applied magnetic field $H$, was varied between 0 and 51.7 kA m$^{-1}$,
resonance was measured for $H$ with and against the bias direction.
The determined resonance field is referred to as $H_{f}$. Microwave
frequencies were chosen such that the fundamental resonance mode
(FMR) and the first exchange mode (FEX) were excited at applied fields
larger then the saturation field of 1.6 kA m$^{-1}$.

\begin{figure}
\begin{centering}
\includegraphics[width=8cm]{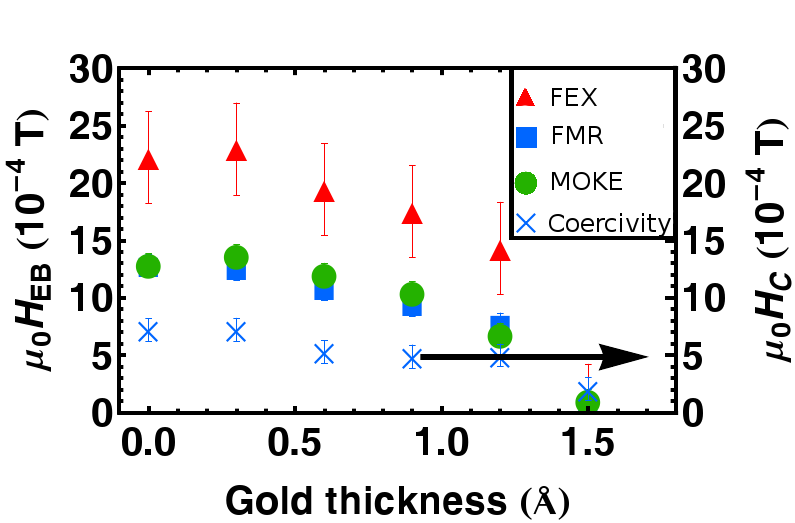}
\par\end{centering}

\caption{\label{fig:bias1}Exchange bias field $H_{EB}$ vs gold dusting thickness
for a NiFe(60.5 nm)/Au(x nm)/IrMn(6 nm) film as measured from the FMR
mode (solid squares), FEX mode (solid triangles) and MOKE (solid circles).
Also shown is the coercive field $H_{C}$ as measured by the MOKE technique with
a 10Hz repetition rate (crosses).}

\end{figure}

As in previous studies\cite{PhysRevB.83.054405}, the bias was
characterised using spin wave modes as half the difference
between $H_{f}$ along and against the easy axis direction. The exchange
bias determined by the two spin wave modes and MOKE are shown
in Fig.(\ref{fig:bias1}). As expected\cite{Measure_Bias_MOM_FMR}
the FMR and MOKE data are in good agreement, and the FEX mode shows
a much larger exchange bias field. The FEX bias shift is larger then the other measures of exchange bias since this mode is particularly sensitive to interface pinning\cite{PhysRevB.83.054405}. Addition
of gold at the interface decreases the exchange bias to zero at an nominal dusting of 1.5 $\textrm{\AA}$. Similar findings have been reported for other types of dusting materials\cite{PhysRevB.77.134401}. We note that coercivity undergoes a very slow decrease with increasing Au
dusting thickness.

\begin{figure}
\begin{centering}
\includegraphics[width=8cm]{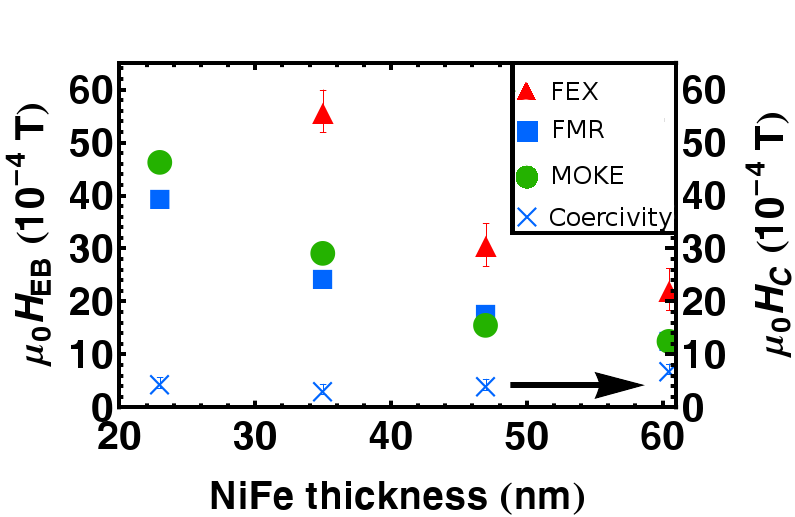}
\par\end{centering}

\caption{\label{fig:bias2}Exchange bias field $H_{EB}$ vs NiFe thickness, of films with no gold dusting and
a 6nm IrMn layer, as measured from the FMR mode (solid squares), FEX
mode (solid triangles) and MOKE (solid circles). Also shown is the
coercive field $H_{C}$ as measured by the MOKE technique with a 10 Hz repetition
rate (crosses).}

\end{figure}

The characteristic inverse ferromagnetic thickness effect for exchange bias $H_{EB}\,\propto\,1/t_{NiFe}$
\cite{Berkowitz1999552} is seen in experiments, shown in Fig.(\ref{fig:bias2}).
Different magnitudes of exchange biasing are seen for different resonance
modes. Coercivity decreases slightly with decreasing permalloy thickness.

\section{Model}

Standing spin wave resonance conditions for in-plane magnetised thin
films can be estimated using Kittel's formula\cite{Kittel_Formula}:
\begin{eqnarray}
\omega & = & \gamma\sqrt{\left(\mu_{0}H_{f}+\mu_{0}M_{s}+Dk^{2}\right)\left(\mu_{0}H_{f}+Dk^{2}\right)}\label{eq:Kittelequation}\end{eqnarray}
 where $\omega$ is the frequency of the spin wave with wavenumber $k$, $H_{f}$ is an externally applied field in the
plane of the thin film, $M_{s}$ is the saturation magnetisation and
$D$ is the exchange constant. In the simplest approximation, perturbations to spin wave frequency
due to interface pinning and surface effects can be associated with a modified $k$. The precision with which FMR can be performed requires a more realistic model however, capable of describing inhomogeneous magnetic parameters near the interface. We base our model on the theory of Benson and Mills\cite{PhysRev.178.839}. This also
describes additional weak surface pinning due to dipole
field effects which are not present in continuous models\footnote{Example frequency shifts are 8.04$\times$10$^{-3}$ GHz for the FMR mode and 24.3$\times$10$^{-3}$ GHz for the FEX mode for a 60.5 nm
permalloy film using calculation parameters $\mu_{0}H_{f}$=$0.02$ T, $\mu_{0}M_{s}$=$0.8$ T and $D$=$1.37\times10^{-17}$ T m$^{2}$.}.

\begin{figure}[h!]
\centering{}\includegraphics[width=6cm]{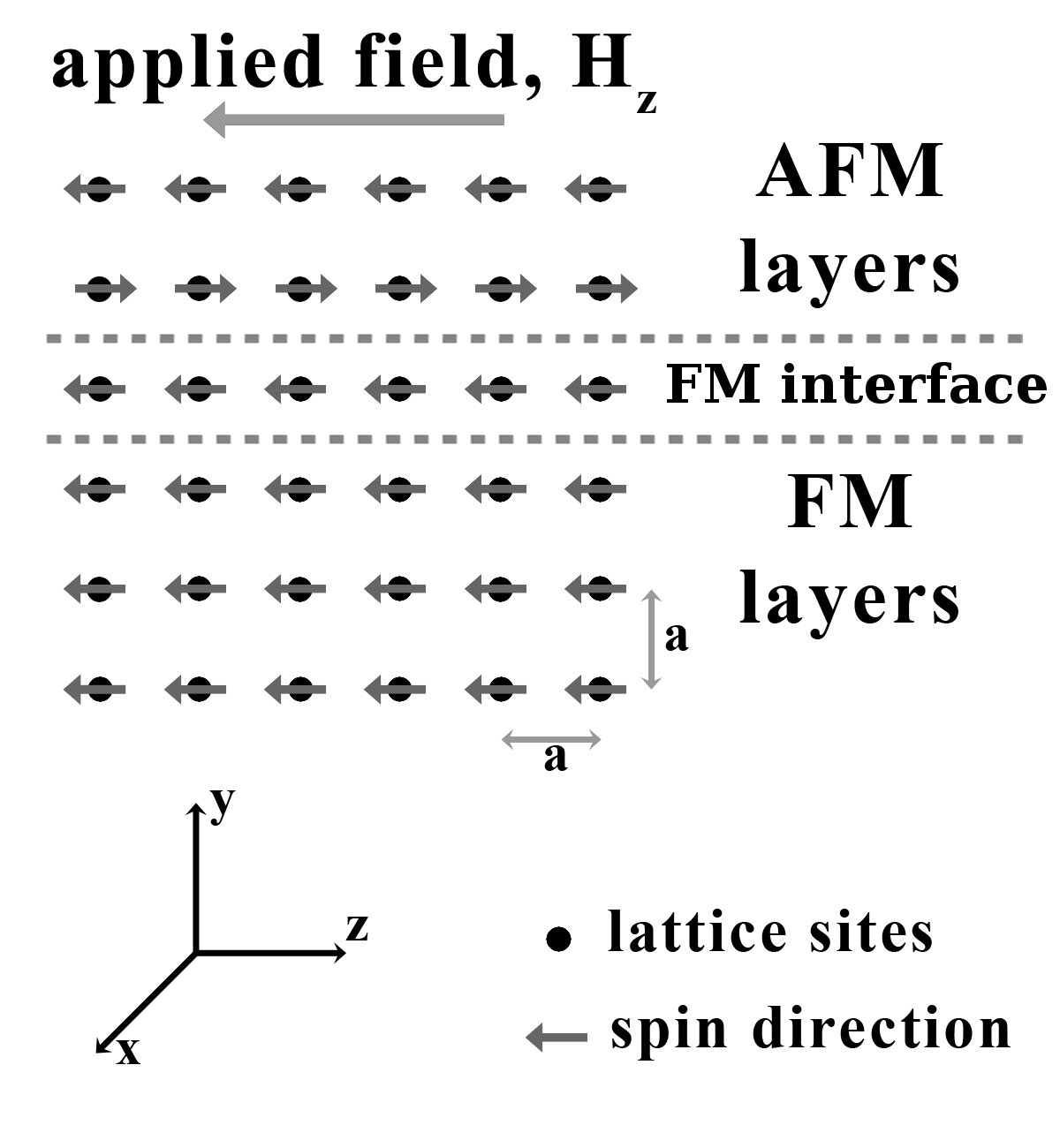}\caption{\label{fig:schematic}Geometry for the atomistic model. The lattice is repeated uniformly in the
out of plane x-direction and is infinite in the x and z directions.
Its structure is simple cubic with lattice constant a.}

\end{figure}

The geometry is shown in Fig.(\ref{fig:schematic}) where there is a discrete number of layers in the y-direction, and an infinite number of lattice sites in the x and z directions. 

The spin direction $S_{z}$, saturation magnetisation $M_{S}$, exchange
interaction $J$, in plane $K_{ip}$ and out of plane $K_{oop}$ anisotropy
are defined individually for each layer.  In what follows, the same set of parameters for all AFM layers is used.  Likewise, the parameters for all FM layers are the same except for the FM interface layer, which can vary. The spins in the ferromagnetic
layers are all aligned along the direction of the externally applied
field and antiferromagnetic layers are aligned antiparallel to neighbouring
antiferromagnetic layers as shown in Fig.(\ref{fig:schematic}). At
the interface the first antiferromagnetic layer can be set either
parallel or antiparallel relative to the applied field and ferromagnetic
layers.  For the purposes of this calculation, interface coupling $J_{int}$ is always set to be positive.  The geometry when the FM and AFM layers are aligned at the interface is referred to as {}``with bias''.  Likewise, the geometry when the FM and AFM layers are anti-aligned at the interface is referred to as {}``against bias''.

Following Ref.\cite{PhysRev.178.839}, there are four contributions
to the Hamiltonian:

\begin{eqnarray*}
\mathcal{H} & = & \mathcal{H}_{zeeman}+\mathcal{H}_{ex}+\mathcal{H}_{dip}+\mathcal{H}_{ani}\end{eqnarray*}

The first contribution is due to the static applied field, $H$, in
the z-direction;

\begin{align}
 & \mathcal{H}_{zeeman}=-g\mu_{B}H_{f}\sum_{\myvec l}S_{z}\left(\myvec l\right)\end{align}
 with $g$ the Land\'{e} factor and $\mu_{B}$ the Bohr magneton. The
second contribution is due the exchange interaction; \begin{align}
 & \mathcal{H}_{ex}=-\frac{1}{2}\sum_{\myvec l}\sum_{\myvec l'\neq\myvec l}J\left(\myvec r\right)\myvec S\left(\myvec l\right)\cdot\myvec S\left(\myvec l'\right)\end{align}
 with $J\left(\myvec l-\myvec l'\right)$ giving the exchange between
two spins at $\myvec l$ and $\myvec l'$, this is only non zero for
nearest neighbours. The third contribution is from dipolar interactions;
\begin{align}
 & \mathcal{H}_{dip}=g^{2}\mu_{B}^{2}\sum_{\myvec l}\sum_{\myvec l'\neq\myvec l}\left\{ \frac{\myvec S\left(\myvec l\right)\cdot\myvec S\left(\myvec l'\right)}{|\myvec r|^{3}}-3\frac{\left[\myvec r\cdot\myvec S\left(\myvec l\right)\right]\left[\myvec r\cdot\myvec S\left(\myvec l'\right)\right]}{|\myvec r|^{5}}\right\} \end{align}
 where $\myvec r=\myvec l-\myvec l'$. This is a long range interaction
and the resulting dipole sums are very slow to converge. Following
the treatment detailed by Ref.\cite{PhysRev.178.839} using Ewald's
method these are converted to a rapidly convergent form.

Further anisotropy contributions, such as magnetocrystalline anisotropies, which arise from sources other then the demagnetizing field comprises both in plane uniaxial anisotropy
\cite{Lévy1993310,CzechPhys.10.1960} in the z direction and
out of plane anisotropy \cite{Kohmoto2003280} in the y direction;
\begin{align}
 & \mathcal{H}_{ani}=-\sum_{\myvec l}K_{ip}\left(\myvec l\right)S_{z}\left(\myvec l\right)^{2}-\sum_{\myvec l}K_{oop}\left(\myvec l\right)S_{y}\left(\myvec l\right)^{2}\end{align}
 where $K_{ip}\left(\myvec l\right)$ is the static in plane anisotropy
constant and $K_{oop}\left(\myvec l\right)$ is the out of plane anisotropy
constant for a spin at $\myvec l$.

Equations of motion are formed from the Hamiltonian by assuming
a translationally invariant plane wave solution and linearising by assuming time invariant terms $S_{z}$. The equations of motion are solved
by expressing them in matrix form and numerically solving the eigenvalue problem
to find the spin wave frequencies and mode profiles. The saturation
magnetisation is defined as $M_{s}=\frac{g\mu_{B}S_{z}}{a^{3}}$,
but in the code $S_{z}$ is used as a factor to alter $M_{s}$.  The link between $D$ and $J$ is given by $\frac{D}{a^2}=\frac{2 J S_{z}}{\mu_{B}}$.

\begin{figure}
\begin{centering}
\includegraphics[width=8cm]{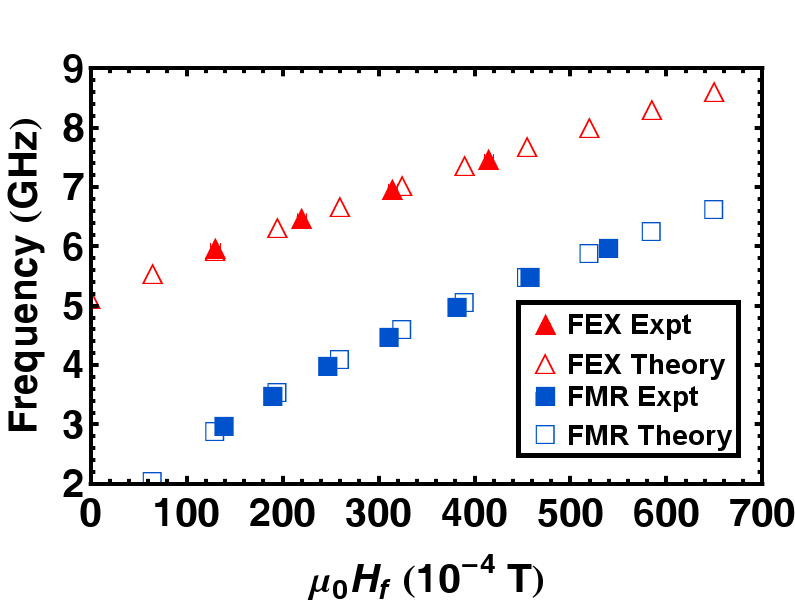}
\par\end{centering}

\caption{\label{fig:wvsH} Experimental and calculated resonances
for the FMR and FEX modes for the 60.5 nm
thick NiFe film. The experiment was performed on the film with a 1.5 $\textrm{\AA}$
gold layer, as it displays no exchange biasing. Filled squares
show the experimental FMR resonances, empty squares show the calculated
FMR resonances, filled triangles show the experimental FEX resonances
and empty triangles show the calculated FEX resonances. Experiment is abbreviated to Expt in the figure legend.}

\end{figure}

Matching excitation frequency $\omega$ vs $H_{f}$ simulation
results for both the FMR and FEX mode to experimental data, as shown
in the example of Fig.(\ref{fig:wvsH}), allows a consistent extraction of bulk in-plane
anisotropy and exchange constant parameters. The sample with the thickest
gold dusting of 1.5 $\textrm{\AA}$ is used to fit $\gamma$, M$_{S}$, K$_{ip,FM}$, K$_{op,FM}$ and $D$ because this sample shows no bias(see Fig.(\ref{fig:bias1})).
Best fits give in-plane anisotropy of the ferromagnet in field units is
$\frac{2\, K{}_{ip,FM}}{M_{S}}=2.5\times10^{-4}$$\pm0.5\times10^{-4}$ T and the spin wave
stiffness is D$=1.48\times10^{-17}$$\pm0.015\times10^{-17}$ T m$^{2}$, which corresponds
to an exchange constant J$_{FM}=5.447\times10^{-22}$ J. Data obtained
by FMR magnetometry is used to set $\gamma=2.8\times 10^{10}$ Hz T$^{-1}$ and M$_{S}=6.3662\times10^{5}$ A m$^{-1}$.
The lattice constant of permalloy was set as a=0.355 nm\cite{123849}
and 170 discrete layers were used to simulate a 60.5 nm thick permalloy film.

Interactions between the antiferromagnet and ferromagnet are mediated
via direct exchange interaction at the FM/AFM interface J$_{int}$,
and long range dipole forces. The above values were held constant and interface parameters varied to fit FMR data from the thinner Au samples.  Parameters used for the antiferromagnet were obtained
from\cite{PhysRevB.79.020403} and are in-plane anisotropy $\frac{2\, K{}_{ip,AFM}}{M_{S}}=2.417$ T
, magnetisation M$_{S,AF}=4.5493\times10^{5}$ A m$^{-1}$, exchange
coupling J$_{AF}=7.67\times10^{-22}$ J and the lattice constant a=0.3785 nm.
16 layers were used in order to simulate a 6 nm thick IrMn film.

\section{Results for Interface Parameters}

\begin{figure}
\centering{}\includegraphics[width=8cm]{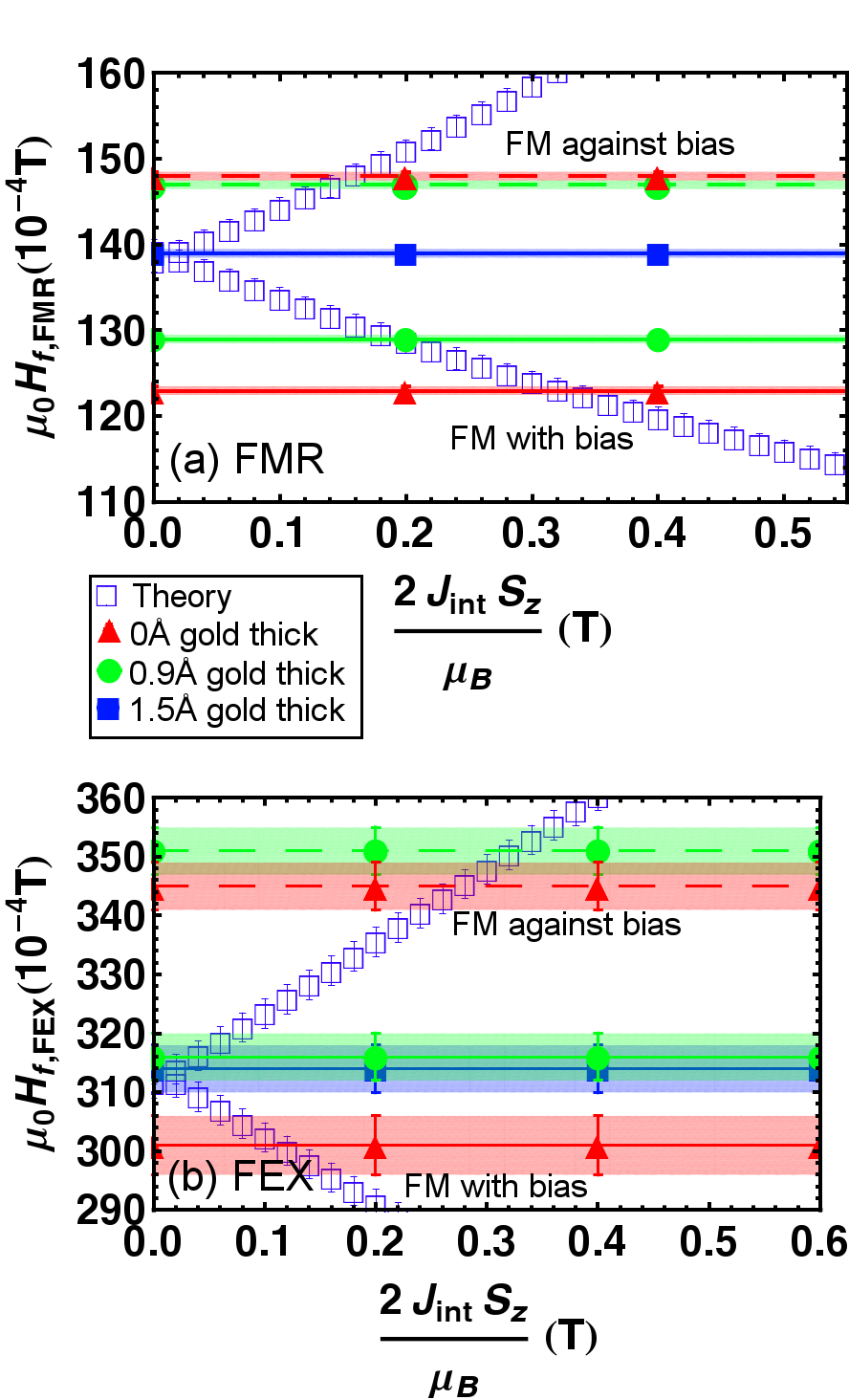}\caption{\label{fig:JintvsH}(a) FMR resonance field for a 60.5 nm NiFe film with a resonance frequency of 3 GHz shown as a function of exchange
coupling J$_{int}$ to a 6 nm thick IrMn antiferromagnet. Empty squares
show calculated resonances when the FM is aligned with bias (represented
by the lower branch) and when the FM is aligned against bias
(represented by the upper branch). For comparison, experimental data
is shown for three dusting thicknesses of gold, 1.5 $\textrm{\AA}$
is the blue (filled squares) line, 0.9 $\textrm{\AA}$ are the green (filled circles)
lines, 0 $\textrm{\AA}$ are the red (filled triangles) lines. Dashed and solid
lines represent against and with bias, respectively. There is only
a single solid line for the 1.5 $\textrm{\AA}$ data as there is no
bias for this sample. (b) The FEX resonance field for a 60.5 nm
NiFe film with a resonance frequency of 7 GHz is shown as a function of exchange
coupling J$_{int}$ to an IrMn antiferromanget. All symbols otherwise
are equivalent to those shown in part (a).  Experimental uncertainties are greater for the resonance field of the FEX mode.}

\end{figure}

We compare the calculated
resonance field $H_{f}$ of the FMR mode (at 3 GHz) and the FEX mode
(at 7 GHz) by altering only the strength of J$_{int}$ in Fig.(\ref{fig:JintvsH}). Experimental data
for Au thicknesses of 0, 0.9 and 1.5 $\textrm{\AA}$ are shown for
comparison. The two different theoretical lines represent the resonance fields for the FM aligned with and against bias. It is these two configurations which produces the observed exchange bias as seen in ferromagnetic resonance experiment.

There are several features of note. Firstly there is a small difference in $H_{f}$ for the two orientations (referred to as a {}``field
gap''), even when J$_{int}=0$. This is caused by the different forces
acting on spins at the interface when the ferromagnet is aligned with
or against the antiferromagnet due to dipolar pinning from the antiferromagnet interface.
Such a feature should not be seen experimentally as the interface is not atomically
flat, and the polycrystalline nature of the thin film introduces
a mixture of interface conditions. Furthermore, we assume that J$_{int}$ and all other interface parameters represent averages across interface.

The field gap in Fig.(\ref{fig:JintvsH}) for the two directions at nonzero J$_{int}$ can
reproduce the exchange bias seen in the experimental data, but only if J$_{int}$ is different for each field direction.  This is difficult to understand and we interpret it as meaning the model is too simple.

As a minimum complication we introduce a quantity $\frac{M_{int}}{M_{S}}$ which describes
changes in the interface magnetisation of the ferromagnet from that
of the bulk. This is not unreasonable as the interface should be directly
modified by the gold dusting layer. In fact, changes in interface magnetisation have been observed experimentally\cite{PhysRevB.81.212404,bruck:126402,PhysRevLett.95.047201,PhysRevLett.91.017203,PhysRevLett.104.217204},
and have also been used in attempts to theoretically describe exchange
bias\cite{Kiwi2001584,PhysRevB.66.014430}. In this way there are effectively magnetic {}``clusters''
coupled to the ferromagnet and antiferromagnet across the interface.

%
%

\begin{figure}
\begin{centering}
\includegraphics[width=8cm]{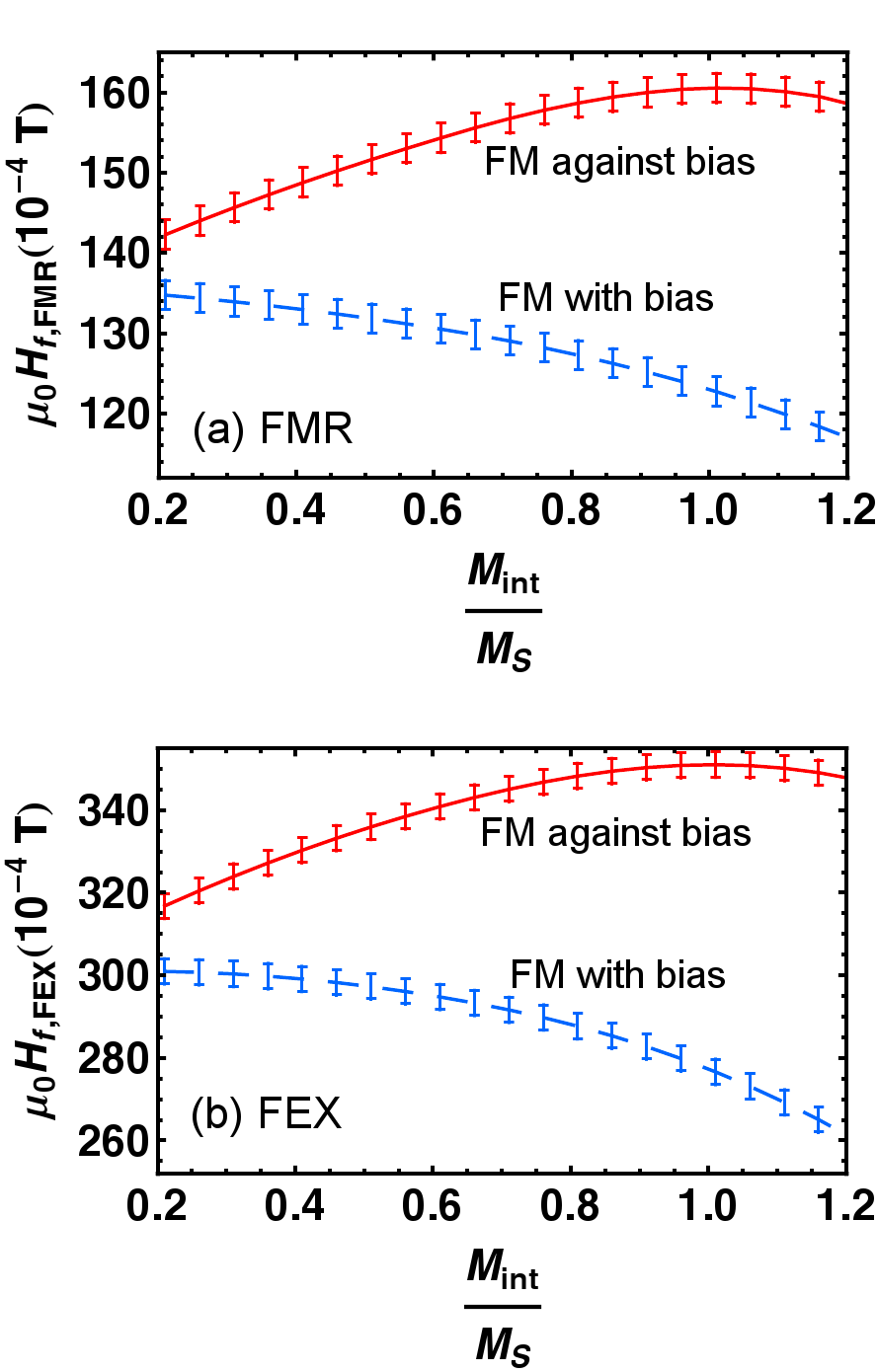}
\par\end{centering}

\caption{\label{fig:dMvsHres} (a) Calculated FMR resonance fields
H$_{f}$ (at a 3 GHz excitation frequency) shown as a function of
interface magnetisation when the ferromagnet is aligned with bias
(dashed line) and against bias (solid line). Likewise (b) shows
how the calculated FEX resonance field H$_{f}$ (at a 7 GHz excitation
frequency) varies as a function of interface magnetisation when the
ferromagnet is with bias(dashed line) and against bias(solid line). Parameters used here are the same as described in
the Model section, with interface coupling set to $\frac{2\, J{}_{int} S_{z}}{\mu_{B}}=0.325$
T.  Ferromagnet is abbreviated as FM for the figure labels.}

\end{figure}

\begin{figure}
\centering{}\includegraphics[width=8cm]{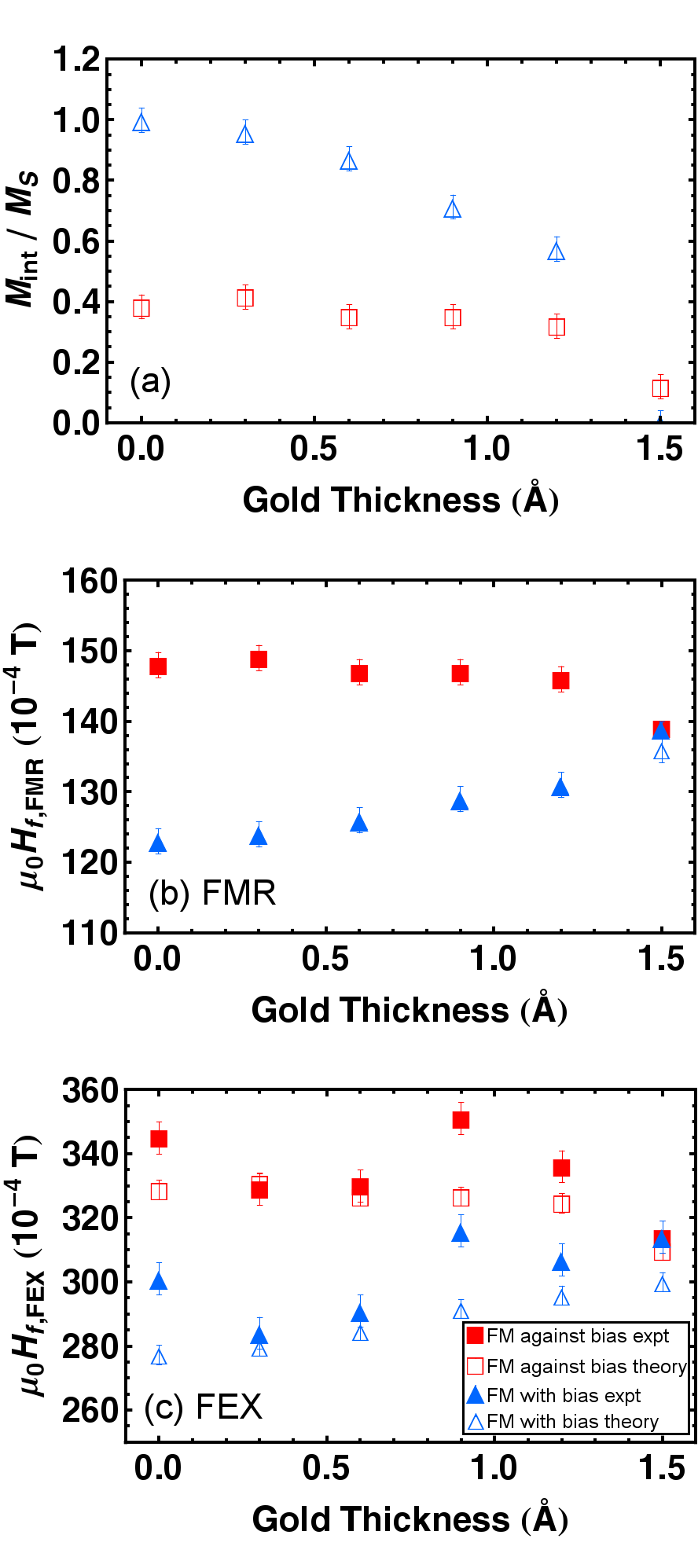}\caption{\label{fig:SimvsExp} Shown are the outcomes of fitting the calculated
data to the FMR modes observed in experiment by allowing the interfacial
magnetisation $\frac{M_{int}}{M_{S}}$ to be different for the two
alignment directions while fixing $\frac{2\mbox{J}_{int} S_{z}}{\mu_{B}}=0.325$
T for both directions. (a) $\frac{M_{int}}{M_{S}}$ is shown for the FM aligned with bias (triangles) and the FM aligned against bias (squares), this is a calculated quantity from the resonance data.
(b) Experimental (solid symbols) and calculated (empty symbols) FMR mode resonance at 3 GHz excitation frequency, fits are sufficiently good that the experimental and theory significantly overlap. (c) Experimental (solid symbols) and calculated (empty symbols) FEX resonance fits at a 7 GHz excitation frequency.  The label for Experiment is abbreviated as expt in the figure legend.}

\end{figure}

The dependence of resonance conditions as a function of $\frac{M_{int}}{M_{S}}$ is shown in Fig.(\ref{fig:dMvsHres})
for a fixed interface coupling to the AFM, set to $\frac{2\, J{}_{int} S_{z}}{\mu_{B}}=0.325$ T. This value of interface
coupling was chosen to correspond to the coupling found for the 0 $\textrm{\AA}$ gold
dusting film, as in Fig.(\ref{fig:SimvsExp})(a).

Both orientations of the FM with respect to the AFM are shown in Fig.(\ref{fig:dMvsHres}). When
the FM is aligned with bias, we note that the resonance field is always decreased
as expected from arguments presented in\cite{PhysRevB.83.054405}. The resonance
field for the FM aligned against bias is always bigger, and has a maximum
before it starts to converge towards the {}``with bias'' case at low $\frac{M_{int}}{M_{S}}$.
This represents the FM and AFM becoming decoupled as the interface
magnetisation decreases to zero. In other words there are two solutions
of a suitable $\frac{M_{int}}{M_{S}}$ value for a given resonance
field for the {}``against bias'' case. In fitting to the data, the lower
$\frac{M_{int}}{M_{S}}$ solution was chosen because $\frac{M_{int}}{M_{S}}$
should decrease as more Au is introduced to the interface.

In order to match calculated resonances to experimental resonances,
$\frac{M_{int}}{M_{S}}$ is allowed to vary depending on FM direction
with respect to the AFM, but J$_{int}$ is required to be the same
regardless of FM orientation. Due to the smaller experimental uncertainties related
to the FMR mode data, the parameters determined this way represent a best fit to optimise agreement with the FMR mode resonance conditions. 

\begin{figure}
\begin{centering}
\includegraphics[width=8cm]{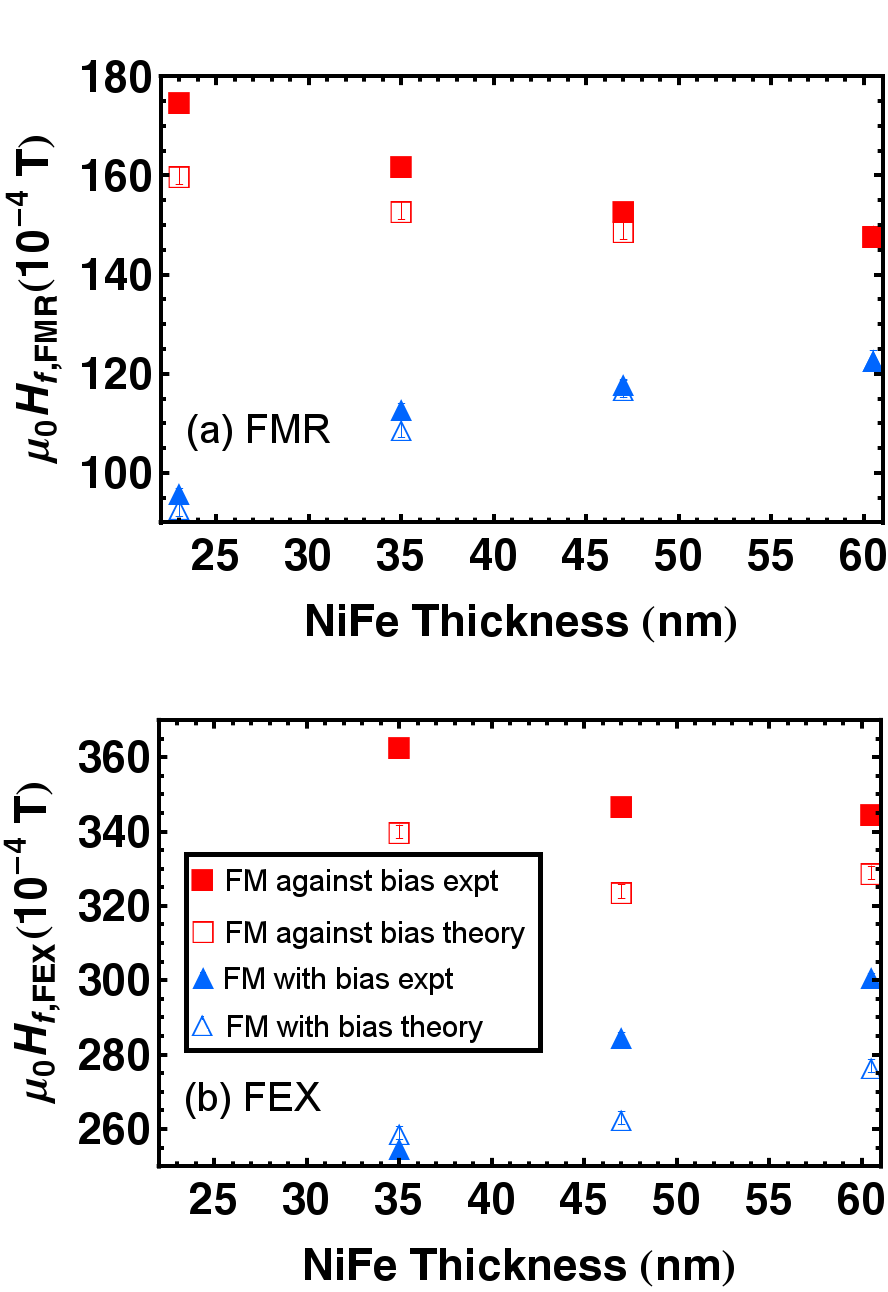}
\par\end{centering}

\caption{\label{fig:thickvsHres}(a) FMR resonance fields (at a 3GHz
excitation frequency) shown as a function of NiFe thickness when the ferromagnet
is with the bias direction (triangles) and against the bias direction
(squares). Experimental results are the solid symbols and theoretical
results are the empty symbols. The relevant simulation parameters
for all results here are $\frac{2\, J{}_{int} S_{z}}{\mu_{B}}=0.325$ T,
$\frac{M_{int}}{M_{S}}$=1 (FM with bias direction) and $\frac{M_{int}}{M_{S}}$=0.383
(FM against bias direction). (b) FMR resonance fields (at
a 7, 8.3, 10.6 GHz excitation frequency for the 60.5, 47, 25nm NiFe
thicknesses respectively) shown as a function of NiFe thickness when the ferromagnet
is with the bias direction (triangles) and against the bias direction
(squares). Experimental results are the solid symbols and theoretical
results are the empty symbols. The relevant simulation parameters
for all results here are the same as in part (a).  The label for Experiment is abbreviated as expt in the figure legend.}

\end{figure}

Fits to J$_{int}$ were done by matching to experimental data for
the 0 $\textrm{\AA}$ gold film and finding $\frac{M_{int}}{M_{S}}\leq1$ solutions.
Fig.(\ref{fig:thickvsHres}) shows the experimental results for the
FMR resonances (at an excitation frequency of 3 GHz) of the FMR mode,
along with simulation results which use J$_{int}$ and $\frac{M_{int}}{M_{S}}$
values found for the Ta(50 $\textrm{\AA}$)/ Ni$_{80}$Fe$_{20}$
(t $\textrm{\AA}$)/Ir$_{25}$Mn$_{75}$ ( 60 $\textrm{\AA}$)/ Ta(50
$\textrm{\AA}$) films. Although there is not complete agreement,
using the parameters found for the thickest of the permalloy films
seems to qualitatively reproduce the experimental resonance
results. Fig.(\ref{fig:thickvsHres}) also shows a comparison
of experimental to theoretical results for the FEX mode in thinner
NiFe films, although one should note the experimental excitation frequencies
are varied from film to film in order to keep the modes in a H$_{f}$
range observable by our setup due to thickness effects. 

Results for $\frac{M_{int}}{M_{S}}$ with $\frac{2\mbox{J}_{int} S_{z}}{\mu_{B}}=0.325$
T, and comparisons to the experimental $H_{f}$ data for both modes
are shown in Fig.(\ref{fig:SimvsExp}).
\clearpage

The calculations are in excellent agreement with the experimental data
for the FMR mode. In contrast, the match to the FEX mode is not as good,
with some large fluctuations which are possibly due to sample-to-sample
thickness variation, as exchange modes are extremely sensitive to
film thickness\cite{Kittel_Formula}. Nevertheless, several points
do closely follow the calculated trend for resonance field shifts.
Interface magnetisation $\frac{M_{int}}{M_{S}}$ is suppressed
when the FM is against bias, but approaches the bulk
value it is aligned with bias. The average $\frac{M_{int}}{M_{S}}$ decreases roughly linearly as a function of gold dusting thickness and results from decreased average exchange coupling across the interface. At a average gold thickness of 1.5 $\textrm{\AA}$
interface magnetisation drops sharply for both orientations as coupling
through the interface layer becomes negligible.

\section{Discussion and Summary}

We have used measurements of standing spin wave modes in a modified
interface exchange biased system to propose a model of effective exchange coupling in an FM/AFM system with diluted interface. Interface coupling mediated via the interface magnetisation is found
to slowly attenuate as the gold dusting thickness was increased, and
then drops to zero at 1.5 $\textrm{\AA}$. When the FM and AFM are
aligned (FM is aligned with bias) and there is no gold dusting, the interface magnetisation is found to be close to the bulk
$M_{s}$ value for the FM. More importantly, the interface magnetisation
is substantially suppressed when the FM is antiparallel aligned to
the AFM (FM is aligned against bias). The interface magnetisation couples to and pins the ferromagnet,
and it is this difference in pinning, arising from different FM/AFM alignment, which produces the bias effect
observed with spin wave resonance. We have shown that a modified interface magnetisation model accurately reproduces the measured resonance fields.

Although modification of antiferromagnetic anisotropy in the model was tried,
it was not sufficient to realistically change the calculated resonances. Resonances in the ferromagnet are found to be  insensitive to changes
in AFM anisotropy for a large range about the standard value\cite{PhysRevB.79.020403} of $\frac{2\, K{}_{ip,AFM}}{M_{S}}=2.417$
T for IrMn. Restricting any changes in anisotropy to the interface layers of
the FM or AFM to reproduce these experimental field shifts leads to
physically unrealistic values. The conclusion is that $\frac{M_{int}}{M_{S}}$
is the most sensible parameter to change in order to describe the
experimental results.

Experimentally it has been found that exchange biased multilayers
have pinned or uncompensated spins\cite{PhysRevB.81.212404,bruck:126402,PhysRevLett.95.047201,PhysRevLett.91.017203,PhysRevLett.104.217204}
and fanning of magnetisation in the ferromagnet\cite{PhysRevB.70.224414}.
Our findings show that the interface magnetisation might vary as the
bulk ferromagnetic spin direction is altered with respect to the antiferromagnet.
The maximum difference of interface magnetisation between the two
FM orientations is 0.6 times the value for bulk $M_{S}$ and is confined
close to the interface. Furthermore, exchange coupling strength across
the ferromagnet/antiferromagnet interface needs only be very small
compared to exchange coupling strength in the bulk of each material
respectively, although we note that this is an effective coupling
as mediated via the layer of altered magnetisation at the interface.
This seems to be true even for samples where the permalloy layer is
made thinner. Introduction of different types of dusting layers at different distances from the interface, and analysis with spin wave techniques would allow further exploration of the local changes to magnetisation at the FM/AFM interface.

\section{Acknowledgments}
Support from the Australian Research Council under the Discovery and
Australian Postgraduate Award programmes is acknowledged. Furthermore,
support from the United Kingdom\textquoteright{}s Engineering and
Physical Sciences Research Council is also acknowledged.



%

\end{document}